\documentclass{epl}

\usepackage{bm}  
\usepackage{graphicx}
\usepackage{amssymb}
\usepackage{epstopdf}
\usepackage{color}
\usepackage{amscd}
\usepackage{epsfig}
\newcommand{\bfr}{{\bf r}}

\newcommand{\bfk}{{\bf k}}
\newcommand{\bfp}{{\bf p}}

\newcommand{\bxi}{{\bm \xi}}

\newcommand{\talpha}{{\tilde{\alpha}}}
\newcommand{\tbeta}{{\tilde{\beta}}}
\newcommand{\trho}{{\tilde{\rho}}}

\newcommand{\nhat}{\hat{n}}
\newcommand{\nvec}{{\bf \hat{n}}}

\title{Bridging the microscopic and the hydrodynamic in active filament solutions}
\author{T. B. Liverpool\inst{1,2} \and M. C. Marchetti\inst{3} }
\institute{
  \inst{1} Department of Applied Mathematics, University of Leeds, Woodhouse Lane, Leeds LS2 9JT, UK\\
  \inst{2} Isaac Newton Institute for Mathematical Sciences, 20 Clarkson Road, Cambridge CB3 0EH,
  UK\\
  \inst{3} Physics Department, Syracuse University, Syracuse, NY 13244, USA
} 

\begin{document}

\maketitle

\begin{abstract}
Hydrodynamic equations for an isotropic solution of active polar
filaments are derived from a microscopic mean-field model of the forces
exchanged between motors and filaments.
We find that a spatial
dependence of the motor stepping rate along the filament is
essential to drive bundle formation. A number of differences arise
as compared to hydrodynamics derived (earlier) from a mesoscopic model
where relative filament velocities were obtained on the basis of
symmetry considerations. Due to the
anisotropy of filament diffusion, motors are capable of generating
net filament motion relative to the solvent. The effect of this
new term on the stability of the homogeneous state is
investigated.
\end{abstract}

\section{Introduction}
Soft active systems are a new and exciting class of complex fluids
to which energy is continuously supplied by  internal or external
sources. Biology provides many  examples  of such systems,
including cell membranes and biopolymer solutions driven by
chemical reactions, living cells moving on a substrate, and the
cytoskeleton of eukariotic cells~\cite{Alberts}. The cytoskeleton
is a complex network of long filamentary proteins (mainly F-actin
and microtubules) cross-linked by  a variety of smaller
proteins~\cite{howard,dogterom95}. Among the latter are clusters
of motor proteins, such as myosin and kinesin,  that use chemical
energy from the hydrolysis of ATP to "walk" along the filaments,
mediating the exchange of forces between
them~\cite{takiguchi,nedelec97,surrey01}.

The self-organization of motor-filament mixtures has been studied
by \emph{in vitro}
experiments~\cite{takiguchi,nedelec97,surrey01}. Complex patterns,
including asters and vortices, have been observed as a function of
motor and ATP concentration in a confined quasi-two-dimensional
geometry~\cite{nedelec97,surrey01}. The high frequency mechanical
response of  active filament solutions has also been studied both
experimentally and theoretically~\cite{humphrey02,TBL_active}. The
study of these simplified model systems should lead to a better
understanding of  the formation and stability of more complex
structures of biological relevance, such as the mitotic spindle
formed during cell division~\cite{Alberts,mogilner04}.

There have been a  number of recent theoretical studies of the
collective dynamics of rigid active filaments. First and most
microscopic, numerical simulations  with detailed modelling of the
filament-motor coupling have been shown to generate patterns
similar to those found in experiments~\cite{nedelec97,surrey01}. A
second interesting development has been the proposal of
'mesoscopic' mean-field kinetic equations where the effect of
motors was incorporated via a motor-induced relative velocity of
pairs of filaments, with the form of such velocity inferred from
general symmetry considerations~\cite{nakazawa96,Kruse00,Kruse01}.
Finally, hydrodynamic equations have been proposed  where the
mixture is described in terms of a few coarse-grained fields whose
dynamics is also inferred from symmetry
considerations~\cite{bassetti00,Lee01,ramaswamy02,Kim03,ramaswamy04,Kruse04,Sankararaman03}.
Recently, we established a connection between the mesoscopic and
hydrodynamic approaches by deriving hydrodynamics via a
coarse-graining of the kinetic equations~\cite{TBLMCM03}. Both the
mesoscopic and hydrodynamic approaches share, however, an
important shortcoming.  The rate and strength of the motor-induced
force exchange among the filaments is controlled by
phenomenological parameters whose dependence on motor activity is
not known. The richness of the phenomena exhibited by the
cytoskeleton leads naturally to the question of how much of the
behavior is specific and how much is generic. To answer this
question it is important to make the connection between
microscopic models and 'generic' hydrodynamic approaches.

In this paper we present a first attempt at deriving the
motor-mediated interaction between filaments from a microscopic
description of the forces exchanged between the motors and the
filaments. Our work establishes the connection between the
hydrodynamic equations and the microscopic motor dynamics. In
particular, it shows explicitly that in mean-field models of the
type considered here spatial inhomogeneities in the motor stepping
rate along the filaments are essential to drive bundle formation.

\section{Kinematics of filament pair}
\begin{figure}
\center \resizebox{0.49\textwidth}{!}{%
  \includegraphics{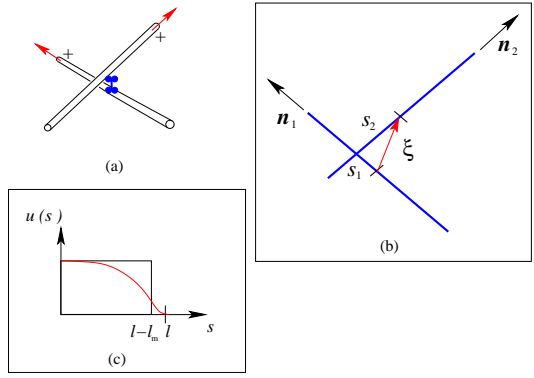}
}

\caption{(a)Two filaments connected by an active crosslink. (b)The
geometry of the overlap: the filaments' centers are separated by
$\bxi=s_1\nvec_1 - s_2\nvec_2$. (c) The profile of the motor
stepping rate.}
\label{2filam}       
\end{figure}

The filaments are modelled as rigid rods of length $l$ (here $l$
should be thought of as the persistence length, rather than the
actual filament length) and diameter $b<<l$. Each filament is
identified by the location ${\bf r}_i$ of its center of mass and a
unit vector ${\bf \hat{n}}_i$ pointing towards the polar end. The
mobile crosslinks are formed by small aggregates of molecular
motors that exert a force on filaments by converting chemical
energy from the hydrolysis of ATP into mechanical work. Each motor
cluster is assumed to be composed of two heads, with the $i$-th
head ($i=1,2$) attached to filament $i$ at position ${\bf
r}_i^{\times}={\bf r}_i+{\bf \hat{n}}_is_i$, with $s_i$ the
position along the filament relative to the center of mass,
$-l/2\leq s_i\leq +l/2$.  The motor cluster has size $l_m<<l$.
Hence the attachment points satisfy ${\bf r}_1^{\times}\simeq{\bf
r}_2^{\times}$, or $\bm{\xi}={\bf r}_2-{\bf r}_1\simeq s_1{\bf
\hat{n}}_1-s_2{\bf \hat{n}}_2$. A schematic is shown in Fig.
\ref{2filam}. Motor heads are assumed to step towards the polar
end of filaments at a known speed, $u(s)$, which generally depends
on the point of attachment. Both filaments and motors move through
a solution. We assume that the filament dynamics is overdamped and
the friction of motors is very small compared to that of
filaments. Momentum conservation then requires that in the absence
of external forces and torques, the total force acting on
filaments centered at a given position be balanced by the
frictional force experienced by the filament while moving through
the fluid.

We consider a pair of filaments cross-linked by a single motor
cluster. Due to the action of the motors, filaments 1 and 2
acquire center-of-mass velocities ${\bf v}_1$ and ${\bf v}_2$ and
rotational velocities $\bm\omega_1$ and $\bm\omega_2$ about the
center of mass. We evaluate these velocities in terms of the known
motor stepping rate, $u(s)$, and of the filaments' orientation. In
the absence of external forces, any force or torque generated by
active crosslinks on one of the filaments of the pair is balanced
by an equal and opposite force or torque acting on the other
filament. This requires
\begin{eqnarray}\label{force_bal} & &\zeta_{ij}({\bf
\hat{n}}_1)v_{1j}= - \zeta_{ij}({\bf \hat{n}}_2)v_{2j}\;,\\
& & \label{torque_bal}  \zeta_r\bm{\omega}_1=-
\zeta_{r}\bm{\omega}_2\;,
\end{eqnarray}
where $\zeta_{ij}({\bf
\hat{n}})=\zeta_\parallel\hat{n}_i\hat{n}_j+\zeta_\perp(\delta_{ij}-\hat{n}_i\hat{n}_j)$
is the friction tensor of the rod, with $\zeta_\parallel$ and
$\zeta_\perp$ the longitudinal and transverse friction
coefficients, respectively, and $\zeta_r$ is the rotational
friction.

The $\alpha$-th motor head ($\alpha=1,2$) at position ${\bf
r}_{m\alpha}={\bf r}_\alpha+s_\alpha{\bf \hat{n}}_\alpha$ has
velocity $\dot{\bf r}_{m\alpha}={\bf v}_\alpha+u_0(s_\alpha){\bf
\hat{n}}_\alpha+s_\alpha\bm{\omega}_\alpha\times{\bf
\hat{n}}_\alpha$, where $\dot{s}_\alpha=u_0(s_\alpha)$ and
$\dot{\bf n}_\alpha=\bm{\omega}_\alpha\times{\bf \hat{n}}_\alpha$.
Motors may also rotate relative to the filament at a rate
$\bm{\omega}_{m\alpha}=\bm{\omega}_\alpha+(-1)^{\alpha-1}\dot{\theta}_\alpha(s_\alpha)({\bf
\hat{n}}_1\times{\bf \hat{n}}_2)/|{\bf \hat{n}}_1\times{\bf
\hat{n}}_2|$, with $\theta_\alpha(s_\alpha)$ the angle the
$\alpha$-th motor head makes with the filament to which it is
attached. If the motor heads are rigidly attached to each other we
must have $\dot{\bf r}_{m1}=\dot{\bf r}_{m2}$ and
$\bm{\omega}_{m1}=\bm{\omega}_{m2}$. Since the motors are assumed
to be point-like in size on the scales of interest, we neglect
motor rotation below, i.e., $\dot\theta_\alpha(s_\alpha)=0$, which
then requires $\bm{\omega}_1=\bm{\omega}_2=0$. The anisotropy of
the rods' friction tensor, allows for both relative and net
translation of the filaments induced by the action of the motors.
Using Eqs.~(\ref{force_bal}), we find that the relative velocity
${\bf v}={\bf v}_1-{\bf v}_2$ of the filaments is
\begin{eqnarray}
\label{vel_eq2} {\bf v}=u_0(s_2){\bf \hat{n}}_2-u_0(s_1){\bf
\hat{n}}_1\;.
\end{eqnarray}
Equations (\ref{force_bal}) and (\ref{vel_eq2}) are readily solved
for the filaments velocities,  ${\bf v}_{1,2}=\pm{\bf v}/2+{\bf
V}$. The velocity ${\bf V}=({\bf v}_1+{\bf v}_2)/2$ of the
 center of mass of the pair is given by
 \begin{eqnarray}\label{V}
{\bf V}= & & A(\sigma,{\bf \hat{n}}_1\cdot{\bf
\hat{n}}_2)\Big\{(1-2\sigma)({\bf \hat{n}}_1\cdot{\bf
\hat{n}}_2)\Big[{\bf
 \hat{n}}_2u_0(s_1)+{\bf
 \hat{n}}_1u_0(s_2)\Big]\nonumber\\
 & &-\Big[1-\sigma-\sigma({\bf
\hat{n}}_1\cdot{\bf \hat{n}}_2)^2\Big]\Big[{\bf
 \hat{n}}_1u_0(s_1)+{\bf
 \hat{n}}_2u_0(s_2)\Big]\Big\}\;,
\end{eqnarray}
where $A=(\sigma/2)\big[(1-\sigma)^2-\sigma^2({\bf
\hat{n}}_1\cdot{\bf \hat{n}}_2)^2\big]^{-1}$, with
$\sigma=(\zeta_\perp-\zeta_\parallel)/(2\zeta_\perp)$. The fact
that ${\bf V}\not=0$ indicates that motor activity can induce a
net motion of the pair relative to the solution. This arises from
hydrodynamic effects due to the anisotropy of the friction tensor
and vanishes when $\zeta_\perp=\zeta_\parallel$. Also ${\bf V}$
vanishes identically for ${\bf \hat{n}}_2=\pm{\bf \hat{n}}_1$, so
that ${\bf V}=0$ in one dimension.


\section{Derivation of hydrodynamics}
The concentration $c({\bf r},{\bf \hat{n}},t)$ of filaments with
center of mass at ${\bf r}$ and orientation ${\bf \hat{n}}$ at
time $t$ satisfies a local conservation law,
\begin{equation}
\partial_t c=-\bm{\nabla}\cdot{\bf J} - \mathcal{\bm R} \cdot \mathcal{\bm
J}\;,
\end{equation}
with $\mathcal{\bm R}={\bf \hat{n}}\times\bm{\nabla}_{\bf
\hat{n}}$. The {\em translational} current density, ${\bf J}$, and
{\em rotational} current density, $\mathcal{\bm J}$, are
\begin{eqnarray}
\label{current}
J_i&=&-D_{ij}\nabla_{j}c-\frac{D_{ij}}{k_BT}c~\nabla_{j}V_{\rm
ex}+J_i^{\rm act}\;, \\
{\cal J}_i &=& -D_r {\cal R}_i c - {D_r \over k_B T} c
{\mathcal{R}}_iV_{\rm ex}
\end{eqnarray}
where
$D_{ij}=k_BT_{act}\zeta^{-1}_{ij}=D_\parallel\nhat_i\nhat_j+D_\perp\big(\delta_{ij}-\nhat_i\nhat_j\big)$
is the translational diffusion tensor, with
$D_{\parallel,\perp}=k_BT_{act}/\zeta_{\parallel,\perp}$ and
$D_r=k_BT_{act}/\zeta_r$.
The effective {\em active} temperature, $T_{act}$ may be higher than the ambient
temperature $T$ due to the motor activity~\cite{TBL_active}.
The potential $V_{\rm ex}$ incorporates
excluded volume effects which give rise to the nematic-transition
in a hard-rod solution. It can be written as $k_BT$ times the
probability of finding another rod within the interaction area of
a given rod, \cite{DoiEdwards}
\begin{eqnarray}
\label{Vex} V_{\rm ex}({\bf r}_1,{\bf
\hat{n}}_1)=\frac{k_BT}{b^d}\int d{\bf r}_2\int d{{\bf
\hat{n}}_2}\int_{s_1}\int_{s_2}\delta({\bf r}_1^{\times}-{\bf
r}_2^{\times})c({\bf r}_2,{\bf \hat{n}}_2,t)\, \vartheta\left( {\bf \hat{n}}_1,{\bf \hat{n}}_2\right)\;,
\end{eqnarray}
where $\int_s...\equiv b^{d-1}\int_{-l/2}^{l/2}~ds...$ and $\displaystyle \vartheta\left( {\bf \hat{n}}_1,{\bf \hat{n}}_2\right) = \sqrt{1-\left(  {\bf \hat{n}}_1 \cdot {\bf \hat{n}}_2 \right)^2}$.

The active current of filaments with center of mass at ${\bf r}_1$
and orientation along $\nvec_1$ is
\begin{eqnarray}
\label{Jact0} {\bf J}^{\rm act}({\bf r}_1,{\bf \hat{n}}_1)=\int
d{\bf r}_2\int d{{\bf \hat{n}}_2}\int_{s_1}\int_{s_2} &&\, \vartheta\left( {\bf \hat{n}}_1,{\bf \hat{n}}_2\right) m({\bf
r}_1^{\times}){\bf v}_1(s_1,s_2,{\bf \hat{n}}_1,{\bf
\hat{n}}_2)\delta({\bf r}_1^{\times}-{\bf r}_2^{\times})\nonumber\\
&&\times c({\bf r}_1,{\bf \hat{n}}_1,t)c({\bf r}_2,{\bf
\hat{n}}_2,t)\;,
\end{eqnarray}
where $m({\bf r})$ is the density of motor clusters, evaluated at
the point of attachment. Finally, ${\bf v}_1(s_1,s_2,{\bf
\hat{n}}_1,{\bf \hat{n}}_2)$ is the velocity that filament 1
acquires at the point of attachment of the motor cluster due to
interaction with filament 2, when the centers of mass of the two
filaments are separated by $\bm{\xi}={\bf r}_2-{\bf
r}_1\simeq\nvec_1s_1-\nvec_2s_2$, as obtained earlier. Using the
$\delta$-function to carry out the integration over ${\bf r}_2$,
and assuming a uniform  density of motor clusters $m_0$, we obtain
\begin{eqnarray}
\label{Jact0} {\bf J}^{\rm act}({\bf r}_1,{\bf \hat{n}}_1)=m_0\int
d{{\bf \hat{n}}_2}\int_{s_1}\int_{s_2}\, \vartheta\left( {\bf \hat{n}}_1,{\bf \hat{n}}_2\right)&&{\bf
v}_1(s_1,s_2,{\bf \hat{n}}_1,{\bf \hat{n}}_2) c({\bf r}_1,{\bf \hat{n}}_1,t)\nonumber\\
&&\times c({\bf r}_1+\nvec_1s_1-\nvec_2s_2,{\bf \hat{n}}_2,t)\;.
\end{eqnarray}
The effect of fluctuations in $m({\bf r})$ will be discussed
elsewhere.

To describe filament dynamics on scales large compared to their
length $l$, we expand the concentration of filaments $c({\bf
r}_1+\bm\xi,{\bf \hat{n}}_2)$ near its value at ${\bf r}_1$. After
inserting this expansion in Eq.~(\ref{Jact0}), the integrals over
$s_1$ and $s_2$ can be carried out term by term and we find
\begin{eqnarray}
\label{Jact0_exp} J^{\rm act}_i({\bf r}_1,{\bf \hat{n}}_1)=m_0\int
d{{\bf \hat{n}}_2}&&c({\bf r}_1,{\bf \hat{n}}_1,t)\Big\{ \langle
v_{1i}\rangle_{s_1,s_2}+\langle
v_{1i}\xi_j\rangle_{s_1,s_2}\nabla_{1j}\nonumber\\
 &&+\frac{1}{2}\langle
v_{1i}\xi_j\xi_k\rangle_{s_1,s_2}\nabla_{1j}\nabla_{1k}+...\Big\}c({\bf
r}_1,{\bf \hat{n}}_2,t)\;,
\end{eqnarray}
where $\langle
v_{1i}\xi_j\xi_k...\rangle_{s_1,s_2}=\int_{s_1}\int_{s_2} \,
\vartheta v_{1i}\xi_j\xi_k...$ are moments of the local filament
velocity. These are expressed in terms of moments of the stepping
speed, $u^{(n)}= \int_{-l/2}^{l/2}(ds/l)(s/l)^n u(s)$. If $u(s)$
is constant, all odd moments, which control the bundling
instability of the homogeneous state discussed earlier
\cite{Kruse00,TBLMCM03}, vanish. In mean-field models of the type
considered here an inhomogeneous velocity profile is crucial to
obtain bundling. Spatial variations of $u(s)$ may for instance
arise from motors slowing down as they approach the polar end of
the filament due to crowding. This is incorporated here by using
the step-like speed profile shown in Fig.~\ref{2filam}, where
$u(s)$ is constant along the filament, but vanishes in a small
region of extent $l_m<<l$ at the polar end. Using this form, we
estimate $u^{(2n)}\sim u_0$ and $u^{(2n+1)}\sim -u_0(l_m/l)$. All
odd moments are negative as filaments slow down as they approach
the polar end.

As in Ref. \cite{TBLMCM03} we now assume that on large scales the
system dynamics can be described in terms of a local filament
density $\rho({\bf r},t)$ and a local filament polarization, ${\bf
p}({\bf r},t)$, defined as the first two moments of the
distribution $c({\bf r},{\bf \hat{n}},t)$,
\begin{eqnarray}
\label{moments} & & \rho({\bf r},t)=\int_{\bf \hat{n}}c({\bf
r},{\bf \hat{n}},t)\;,\nonumber\\
& & {\bf p}({\bf r},t)=\int_{\bf \hat{n}}{\bf \hat{n}}~c({\bf
r},{\bf \hat{n}},t)\;.
\end{eqnarray}
Hydrodynamic equations for these coarse-grained densities can then
be obtained by writing an exact moment expansion for  $c({\bf
r},{\bf \hat{n}},t)$ and truncating it at the second moment. The
integrals over the unit vectors ${\bf \hat{n}}_1$ and ${\bf
\hat{n}}_2$ are carried out by approximating the prefactor $A$ of
Eq.~(\ref{V}) and the excluded volume term $\displaystyle
\vartheta$ with their mean value over the range of integration.
This approximation does not affect the structure of the
hydrodynamic equations, but only the numerical values of the
coefficients.

The nonlinear hydrodynamic equations for the filament density and
local orientation of a mixture of arbitrarily-shaped filaments and
motors in $d$ dimensions will be given elsewhere. Here we restrict
ourselves to the specific case of long thin rods in $d=2$, where
$D_\parallel=2D_\perp\equiv D$. We also linearize the equations
around the homogeneous isotropic state, by letting
$\rho=\rho_0+\delta\rho$ and ${\bf p}=\delta{\bf p}$. The
linearized equations in $d=2$ are
\begin{eqnarray}
\partial_t\delta\rho=&&\frac{3}{4}D\nabla^2\delta\rho-\frac{\alpha\tilde{\rho}_0}{2}\nabla^2\delta\rho
+\frac{\beta\tilde{\rho}_0}{4\sqrt{2}}\bm{\nabla}\cdot{\bf p}\nonumber\\
&& -\frac{5\beta\tilde{\rho}_0l^2}{192}\left(
1-\frac{5}{24\sqrt{2}}\right)\nabla^2\bm{\nabla}\cdot{\bf p} -{5
\alpha \trho_0 l^2\over 384}  \nabla^2 \nabla^2 \delta \rho \;,
\label{rho_lineq}\end{eqnarray}
\begin{eqnarray}
\partial_t p_i=& &-D_r \, p_i+\frac{5}{8}D\nabla^2 p_i+\frac{D}{4}\nabla_{i}\bm{\nabla}\cdot{\bf p}
   +\frac{\beta\tilde{\rho}_0}{2}\Big(1+\frac{1}{4\sqrt{2}}\Big)\nabla_{i}\delta\rho\nonumber\\
   & &+\frac{5\beta\tilde{\rho}_0 l^2}{384}\Big(1-\frac{5}{24\sqrt{2}}\Big)\nabla_{i}\nabla^2\delta\rho\;,
\label{p_lineq}\end{eqnarray}
where $\tilde{\rho}_0=\rho_0v_0$ and
\begin{eqnarray}
& & \alpha=-m_0v_0u^{(1)}\approx m_0v_0u_0l_m/ \pi \;,\\
& & \beta=m_0v_0u^{(0)}=2m_0v_0u_0/ \pi \;.
\end{eqnarray}
The parameter $\alpha$ has the dimensions of a diffusion constant
and describes filament bunching or bundling, which,
 in contrast to conventional diffusion, tends to enhance density fluctuations. The coefficient $\beta$ is a velocity
 and describes the rate at which motor clusters sort or separate filaments of opposite polarity.
If the motor stepping speed $u(s)$ is constant, independent of the
position $s$ along the filament, then $\alpha=0$. In general, even
when $\alpha\not=0$, we expect $\alpha<<\beta$.

\begin{figure}
\center \resizebox{0.49\textwidth}{!}{%
  \includegraphics{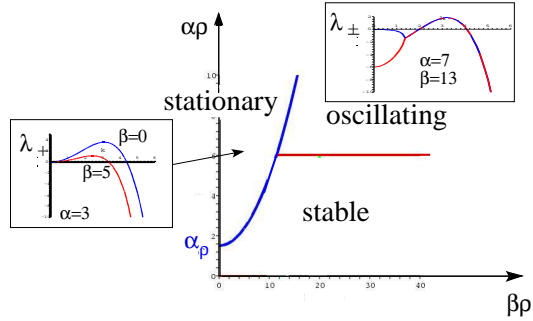}
}

\caption{The phase diagram in the $\talpha-\tbeta$ plane showing
the different regimes calculated from a low $k$ expansion of
$\lambda_+$ and the eigen-vector $e_+^{L,R}$ for $\trho_0 = 0.9$.
The insets show the unstable modes.}
\label{avb}       
\end{figure}

The hydrodynamic equations obtained here by using a microscopic
model for the motor-induced filament velocities differ from those
derived in Ref. \cite{TBLMCM03} where these velocities where
written down on the basis of symmetry in two important ways:
\begin{enumerate}
\item The density equation contains a new term
$\sim\beta\bm{\nabla}\cdot(\rho{\bf p})$ that describes filament
convection along the direction of local alignment. This term
vanishes for isotropic objects with $\zeta_\perp=\zeta_\parallel$.
It arises for rods because in this case the motors can generate
net motion of the filaments relative to the solution (${\bf
V}\not=0$). \item The second important difference is that there
are no $\alpha$-type contributions to the active currents
proportional to gradients of the local polarization. These leads
to the absence of terms of the form $\sim
\nabla_{i}[p_j\nabla_{j}p_i+...]$ in the density equation and of
terms $\sim\nabla_{j}\rho[\nabla_{j}p_i+...]$ in the polarization
equation. These terms are allowed by symmetry and are indeed
present in hydrodynamic models written down on the basis of
symmetry consideration.\cite{Kruse00,TBLMCM03} They vanish,
however, upon coarse-graining, when averaging over the length of
the filaments. The absence of such terms does not change the
bundling instability of density fluctuations, but it does
eliminate the possibility of the orientational instability of
polarization fluctuations observed in \cite{TBLMCM03} at high
filament concentrations.
\end{enumerate}
The absence of terms proportional to polarization gradients in the
active current is not related to the fact that our model does not
incorporate any motor-induced filament rotation.  It arises
instead from the fact that the motor dynamics is described in
terms  of the rate $u(s)$ at which each motor head steps on each
filament. These terms then average out when  the force exchanged
between the two filaments of a pair is averaged along the length
of each filament.

\section{Stability Analysis}

To determine the linear stability of the isotropic, homogeneous
state, we expand density and polarization fluctuations in Fourier
modes, $\delta \rho (\bfr) = \int_k e^{i {\bf k \cdot r}} \rho_k,
\bfp(\bfr) = \int_k e^{i {\bf k \cdot r}} \bfp_k$. Defining the
longitudinal polarization mode $p_k^L = \hat{\bfk} \cdot \bfp_k$
and $\vec{V} = (\delta \rho, p_k^L)$, Eqns.
(\ref{rho_lineq},\ref{p_lineq}) are equivalent to a matrix
equation $\partial_t \vec{V} = {\Bbb M} \cdot \vec{V} \quad $. The
instability of the system to density and longitudinal polarization
fluctuations is determined the range of parameters where for some
range of $\bfk$-vectors, the largest of the eigenvalues
$\lambda_\pm(\bfk)$ of the matrix ${\Bbb
M}(k,\trho_0,\talpha,\tbeta)$ has a positive real part. Expressing
lengths in units of $l$ and time in units of $l^2/D$ we define two
dimensionless parameters $\talpha=\alpha/D$ and $\tbeta=l\beta /D$
\cite{foot_parameters}. For $\tbeta=0$ the equations are
decoupled. In this case there is a stationary instability of
density fluctuations for
$\talpha>\talpha_\rho^0=3/(2\tilde{\rho}_0)$ and $0<k<k_\alpha^0$,
with $k_\alpha^0\sim(\talpha-\talpha_c^0)^{1/2}$. A finite, but
small value of the sorting rate $\tbeta$ tends to stabilize the
homogeneous state. This is easily seen from the linearized
equations as for times $ \gg D_r^{-1}$ longitudinal polarization
fluctuations simply follow density fluctuations, and to leading
order in the gradients $p_k^L\sim
[\beta\tilde{\rho}_0/(2D_r)]ik\rho_k$. Eliminating the
longitudinal polarization in the equation for the density, we find
that a finite sorting rate $\tbeta$ effectively enhances
diffusion, shifting the instability to
$\talpha_\rho(\tbeta)=3/(2\tilde{\rho}_0)+(\tbeta^2\tilde{\rho}_0^2(1+1/(4\sqrt{2}))/(24\sqrt{2})$.
The range of wavevectors over which the modes are unstable is
again $0\leq k\leq k_\alpha(\tbeta)$, with
$k_\alpha(\tbeta)\sim[\talpha-\talpha_\rho(\tbeta)]^{1/2}$. For
larger values of $\tbeta$ the modes are complex conjugate,
describing propagating fluctuations. The homogeneous state becomes
unstable via an oscillatory instability.  The location of the
instability is still, however, controlled solely by the bundling
coefficient $\talpha$. It occurs at the $\tbeta$-independent
value, $\talpha_{m}\approx 7.6/\tilde{\rho}_0$ above which
$Re(\lambda_\pm(k))$ has a positive maximum, for an intermediate
range of wavenumbers. The actual boundaries of the various regions
have been evaluated  numerically and are shown in Fig. 2. To
establish which order parameter (density, polarization, or both)
goes unstable, we have evaluated the right and left eigen-vectors
associated with the unstable mode, defined by ${\Bbb M} \cdot
\vec{e}^R_\pm(k) = \lambda_\pm(k)\vec{e}^R_\pm(k)$ and
$\vec{e}^L_\pm(k)\cdot{\Bbb M} = \lambda_\pm(k)\vec{e}^L_\pm(k)$,
respectively. The unstable eigen-vector associated with the
stationary  instability at $\talpha_\rho(\tbeta)$ is dominated by
density fluctuations at all length scales. The unstable
eigen-vector associated with the oscillatory instability at
$\talpha_m$ is always a mixed one which crosses over from density
at small $k$ to longitudinal polarization at large $k$.

In summary, the homogeneous state is always driven unstable by a
sufficiently large value of the bundling constant $\alpha$. The
polarization sorting rate $\beta$ tends to stabilize the
homogeneous state. A large $\beta$ also changes the nature of the
instability from stationary to oscillatory. Finally, an important
consequence of the absence of $\sim\alpha$-type contributions to
the active currents proportional to polarization gradients is that
our model yields no pure polarization instability at large length
scales. In particular, fluctuations in the transverse part of the
polarization ($\sim\bm{\nabla}\times{\bf p}$) that may control the
onset of vortex-type structures, are always stable to linear
order.  A numerical solution of the full nonlinear equations is,
however, needed to establish the precise nature of the inhomogeneous
states.

\acknowledgments TBL acknowledges the support of the Royal Society.
MCM acknowledges support from the National
Science Foundation, grant DMR-0305407.


\begin{thebibliography}{0}

\bibitem{Alberts}
\Name{B. Alberts et al} \Book{Molecular Biology of the Cell}
\Publ{Garland, New York} \Year{2002}.


\bibitem{howard}
\Name{J. Howard}  \Book{Mechanics of Motor Proteins and the
Cytoskeleton} \Publ{Sinauer, New York} \Year{2000}.

\bibitem{dogterom95}
\Name{M. Dogterom, A. C. Maggs \and S. Leibler} \REVIEW{Proc. Nat.
Acad. Sci. USA}{92}{1995}{6683}.

\bibitem{takiguchi}
\Name{K. Takiguchi} \REVIEW{J. Biochem.}{109}{1991}{520}; \Name{R.
Urrutia et al.} \REVIEW{PNAS}{88}{1991}{6701}.

\bibitem{nedelec97}
\Name{F. J. N\'ed\'elec, T. Surrey, A. C. Maggs \and S. Leibler}
\REVIEW{Nature}{389}{1997}{305}.

\bibitem{surrey01}
\Name{T. Surrey, F. J. N\'ed\'elec, S. Leibler \and E. Karsenti}
\REVIEW{Science}{292}{2001}{1167}.


\bibitem{humphrey02}
\Name{D. Humphrey, C. Duggan, D. Saha, D. Smith \and J. K\"as}
\REVIEW{Nature}{416}{2002}{413}.

\bibitem{TBL_active}
\Name{T. B. Liverpool, A.C. Maggs \and A. Ajdari}  \REVIEW{Phys.
Rev. Lett.}{86}{2001}{4171}.

\bibitem{mogilner04}
\Name{E. Cytrynbaum, V. Rodionov \and A. Mogilner}
\REVIEW{J. Cell Sci.}{117}{2004} {1381}.



\bibitem{nakazawa96}
\Name{H. Nakazawa \and K. Sekimoto} \REVIEW{J. Phys. Soc. Jpn.}
{65}{1996}{2404}; \Name{K. Sekimoto and H. Nakazawa} in
\Book{Current Topics in Physics}\Editor{Y. M. Cho, J B. Homg and
C. N. Yang}\Publ{World Scientific, Singapore}\Year{1998}.

\bibitem{Kruse00}
\Name{K. Kruse \and F J\"ulicher} \REVIEW{Phys. Rev.
Lett.}{85}{2000}{1778}; \REVIEW{Phys. Rev. E}{67}{2003}{051913}.

\bibitem{Kruse01}
\Name{K. Kruse, S. Camalet \and F. J\"ulicher} \REVIEW{Phys. Rev.
Lett.}{87}{2001}{138101}.

\bibitem{Lee01}
\Name{H. Y. Lee \and M. Kardar} \REVIEW{Phys. Rev.
E}{64}{2001}{56113}.

\bibitem{Kim03}
\Name{J. Kim et al.} \REVIEW{J. Korean Phys. Soc.}
{42}{2003}{162}.

\bibitem{Sankararaman03}
\Name{S. Sankararaman, G. I. Menon \and P.B. S. Kumar}
arXiv:cond-mat/0307720.

\bibitem{Kruse04}
\Name{K. Kruse et al}  \REVIEW{Phys. Rev. Lett.}{92}{2004}{078101}.

\bibitem{ramaswamy02}
\Name{R. Aditi Simha \and S. Ramaswamy} \REVIEW{Phys. Rev. Lett.}
{89}{2002}{058101}.

\bibitem{ramaswamy04}
\Name{Y. Hatwalne et al.} \REVIEW{Phys. Rev.
Lett.}{92}{2004}{118101}.

\bibitem{bassetti00}
\Name{B. Bassetti, M. C. Lagomarsino \and P. Jona} \REVIEW{Eur.
Phys. J. B}{15}{2000}{483}.

\bibitem{TBLMCM03}
\Name{T. B. Liverpool \and M. C. Marchetti} \REVIEW{Phys. Rev.
Lett.}{90}{2002}{138102}.

\bibitem{DoiEdwards}
\Name{M. Doi \and S. F. Edwards} \Book{The theory of polymer
dynamics} \Publ{Oxford University Press, New York} \Year{1986}.

\bibitem{foot_parameters}
For the specific form of $u(s)$ used here, $\talpha =
\tbeta(l_m/l)$.

\end{thebibliography}
\end{document}